\documentclass[draft,prb,twocolumn,showpacs]{revtex4}

\pdfoutput=1

\usepackage{amsmath,amsfonts,amssymb,bm}
\usepackage{dcolumn}
\usepackage[final]{graphicx}
\usepackage{bm}

\begin{document}

\bibliographystyle{apsrev}

\title{Strong coupling between a metallic nanoparticle and a single molecule}

\author{Andreas Tr\"ugler}
\author{Ulrich Hohenester}\email{ulrich.hohenester@uni-graz.at}
\affiliation{Institut f\"ur Physik,
  Karl--Franzens--Universit\"at Graz, Universit\"atsplatz 5,
  8010 Graz, Austria}

\date{February 12, 2008}

\begin{abstract}

We theoretically investigate strong coupling between a single molecule and a single metallic nanoparticle. A theory suited for the quantum-mechanical description of surface plasmon polaritons (SPPs) is developed. The coupling between these SPPs and a single molecule, and the modified molecular dynamics in presence of the nanoparticle are described within a combined Drude and boundary-element-method approach. Our results show that strong coupling is possible for single molecules and metallic nanoparticles, and can be observed in fluorescence spectroscopy through the splitting of emission peaks.

\end{abstract}



\maketitle

 
\section{Introduction}
 
Quantum optics has recently made its way to the field of plasmonics.~\cite{chang:06,akimov:07} This is due to the the rapid progress in nanofabrication and measurement techniques. Recent experiments have demonstrated the controlled coupling of single molecules with metallic nanoparticles (MNPs)~\cite{anger:06,kuehn:06,gerhardt:07,akimov:07} and metallic surfaces,~\cite{labeau:07} of coupled nanoparticles,~\cite{rechberger:03,danckwerts:07} and of donor and acceptor molecules accross metal films.~\cite{andrew:04} Possible applications of such hybrid molecule-MNP systems range from biosensing~\cite{stuart:05,yin:05} to active plasmonic devices.~\cite{fedutik:07}

A key element of the quantum-optics toolbox is the strong coupling between a quantum emitter and a resonator, where excitation energy is coherently transferred between emitter and resonator. Strong coupling was first observed for single atoms in high-finesse optical resonators,~\cite{turchette:95,raimond:01} and more recently for various solid state systems, such as semiconductor quantum dots \cite{hennessy:07,press:07} or superconductor circuits.~\cite{wallraff:04} Although strong coupling between ensembles of molecules, e.g, $J$-aggregates of dyes, with plasmons has been reported,~\cite{specialissue.organic} it is unclear whether the strong coupling regime can be reached for single molecules coupled to MNPs. The reason for this lies in the intricate interplay of the molecule-MNP coupling strength with the molecular relaxation dynamics, which becomes heavily altered in the vicinity of the nanoparticle.
 
It is the purpose of this paper to theoretically investigate the strong coupling regime between a single quantum emitter, such as a molecule or collodial quantum dot, and a single MNP. We start by developing a theory suited for the quantum-mechanical description of surface plasmon polaritons (SPPs), the coupling between these SPPs and single molecules, and the modified molecular dynamics in presence of the MNP. We employ a Drude framework for the description of the metal dynamics, and compute the quantized SPP modes within a boundary element method approach.~\cite{hohenester.prb:05,gerber.prb:07} Our results show that strong coupling is possible for molecules and MNPs and could be observed in fluorescence spectroscopy through the splitting of emission peaks.

We have organized this paper as follows. In section \ref{sec:theory} we show how to compute surface plasmon modes within a boundary-element-method approach, and introduce a suitable quantization scheme for the surface plasmons. We also present details of the theoretical description of the coupled molecule-MNP system in presence of scatterings. Sec.~\ref{sec:results} presents results of our model calculations. We explore the strong-coupling regime for a single molecule coupled to a MNP, and identify the pertinent parameters for strong coupling. We also discuss limitations of our model. Finally, in Sec.~\ref{sec:conclusions} we summarize and draw some conclusions.


\section{Theory}\label{sec:theory}

\subsection{Plasmon quantization}

Although SPPs are generally considered as bosonic quasiparticles, most theoretical work does not explicitly rely on such description. In linear response one can employ the fluctuation-dissipation theorem to relate the dielectric response to the dyadic Green tensor of Maxwell's theory,~\cite{vogel:06,hohenester.ieee:08} where all details of the metal dynamics are embodied in the dielectric function, which can be obtained from either experiment~\cite{johnson:72} or first principles calculations. This approach is no longer applicable in nonlinear response. Also the neglect of plasmon relaxation at small timescales, as proposed in other work~\cite{bergman:03}, is not suited for the investigation of strong coupling, which critically depends on the relative importance of coupling and SPP dephasing. In this work we thus follow the seminal work of Ritchie,~\cite{ritchie:57} where the electron dynamics in the metal is described within the hydrodynamic model.~\cite{barton:79} For the transition metals Ag and Au, electrons with particle density $n_0$ are assumed to move freely in a medium with background dielectric constant $\epsilon_0$, which accounts for the screening of $d$-band electrons.~\cite{ladstaedter.prb:04,remark.drude}

\begin{figure}
\centerline{\includegraphics[width=0.8\columnwidth]{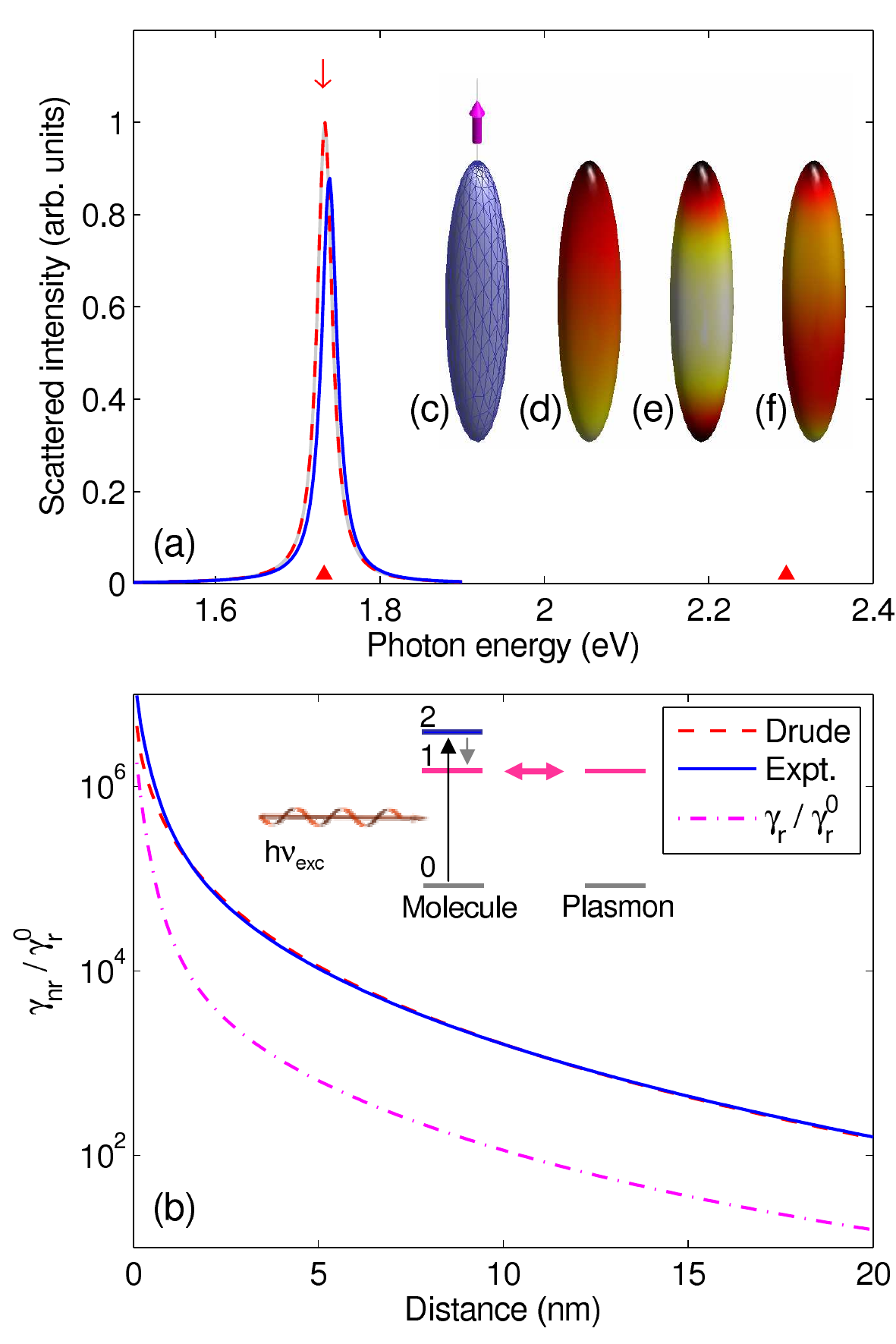}}
\caption{(Color online) (a) Spectrum of cigar-shaped Ag MNP. The height of the particle is 40 nm and the height-to-diameter ratio is $5:1$. The solid and dashed lines show the calculated spectra for the dielectric function of Ref.~\onlinecite{johnson:72} and the Drude form,~\cite{remark.drude} respectively. The triangles at the bottom indicate the energies of the plasmon modes (d) and (e), and the arrow at $\sim 1.7$ eV indicates the energy of the molecular state $1$. (b) Non-radiative (solid line) and radiative (dashed-dotted line) decay rate for a molecule located at a certain distance from the MNP [see panel (c)] in units of the radiative free-space decay rate $\gamma_r^0$. The inset reports the level scheme used in our caluclations. (c) Discretized particle surface as used in our calculations. (d--f) Surface charge distribution of SPP eigenmodes of lowest energy. Only mode (d) has a nonzero dipole moment.
}
\end{figure}

\begin{figure}
\centerline{\includegraphics[width=0.8\columnwidth]{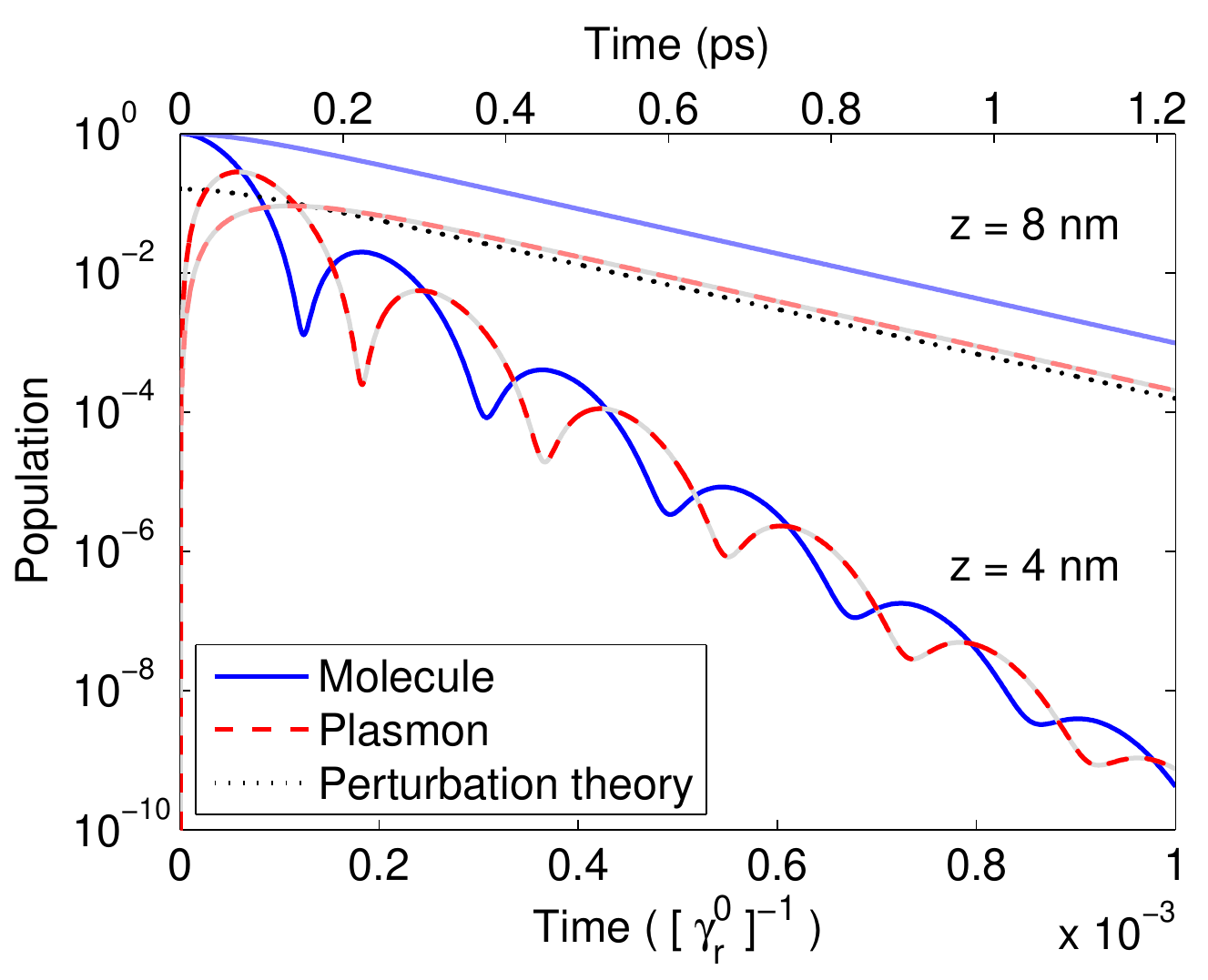}}
\caption{(Color online) Simulation for molecule which is initially in state $1$ [inset of Fig.~1(b)]. The solid and dashed lines show the population of the molecular state $1$ and the SPP dipole mode [Fig.~1(d)], respectively, for two different molecule--MNP distances. We use a molecule dipole moment of 10 atomic units, which corresponds to a free-space decay time $1/\gamma_r^0$ of approximately one nanosecond. In weak coupling (corresponding to the 8 nm distance) the non-radiative decay rate can be also estimated from time-dependent perturbation theory, $\gamma_{nr}\approx |g_\lambda|^2/\gamma_0$, where $g_\lambda$ is the coupling constant between the molecule and the resonant plasmon mode. The dotted line in the figure shows the resulting decay. In our simulations we include the 40 plasmon modes of lowest energy. 
 }
\end{figure}

The energy of a classical electron plasma is the sum of kinetic and electrostatic energy~\cite{ritchie:57,barton:79}
\begin{equation}\label{eq:energyplasma}
  H=\mbox{$\frac 12$}\int\left\{n_0(\nabla\Psi)^2+\rho\Phi\right\}d^3r\,.
\end{equation}
Here $\rho(\bm r)$ is the charge density displacement from equilibrium, $\Phi(\bm r)$ is the electrostatic potential induced by $\rho(\bm r)$, and $\Psi(\bm r)$ is the velocity potential, whose derivative gives the velocity density $\bm v=-\nabla\Psi$.~\cite{barton:79} Throughout we use Gauss and atomic units ($e=m=\hbar=1$). For the SPPs of our present concern we consider surface charge distributions $\sigma$ which are nonzero only at the surface of the MNP. As detailed in Appendix~\ref{app:supplementary}, the Hamilton function \eqref{eq:energyplasma} can be rewritten in a boundary element method (BEM) approach~\cite{hohenester.prb:05,garcia:02} as
\begin{eqnarray}\label{eq:energyspp}
  H&=&\frac 1{2n_0}\Bigl\{\dot\sigma^T\left(2\pi+F\right)^{-1}G\,\dot\sigma
  \nonumber\\&&\quad
  +\omega_p^2\,\sigma^T
  \left[2\pi(\epsilon_0+\epsilon_b)+(\epsilon_0-\epsilon_b)F\right]^{-1}G\,
  \sigma\Bigr\}\quad\,.
\end{eqnarray}
Here $\sigma$ is the vector of the surface charges within the discretized surface elements (see inset of Fig.~1), $G$ is the free Green function matrix which connects two surface elements, $F$ is the corresponding surface derivative,~\cite{hohenester.prb:05,garcia:02} $\omega_p=(4\pi n_0)^{\frac 12}$ is the plasma frequency, $\epsilon_0$ is the background dielectric constant of the metal \cite{remark.drude}, and  $\epsilon_b$ the dielectric constant of the embedding medium. We can now determine the eigenmodes of Eq.~\eqref{eq:energyspp} and quantize the plasma oscillations via a canonical transformation, following the standard procedure outlined in Refs.~\onlinecite{ritchie:57,barton:79,hohenester.ieee:08}. Within such an approach we obtain the plasmon Hamiltonian $H_{\rm pl}=\sum_\lambda \omega_\lambda\,a_\lambda^\dagger a_\lambda^{\phantom\dagger}$ in second-quantized form, with $\omega_\lambda$ being the energy and $a_\lambda^\dagger$ the creation operator for the plasmon mode $\lambda$. The field operator for the SPPs is of the form
\begin{equation}\label{eq:quantspp}
  \sigma(\bm r)=\sum_\lambda \left(\frac{2n_0}{\omega_\lambda\beta_\lambda}\right)^{\frac 12}
  u_\lambda(\bm r)\left(a_\lambda^{\phantom\dagger}+a_\lambda^\dagger\right)\,,
\end{equation}
where $u_\lambda(\bm r)$ is the plasmon eigenfunction and $\beta_\lambda$ the corresponding normalization constant.


\begin{figure*}
\centerline{\includegraphics[width=1.5\columnwidth]{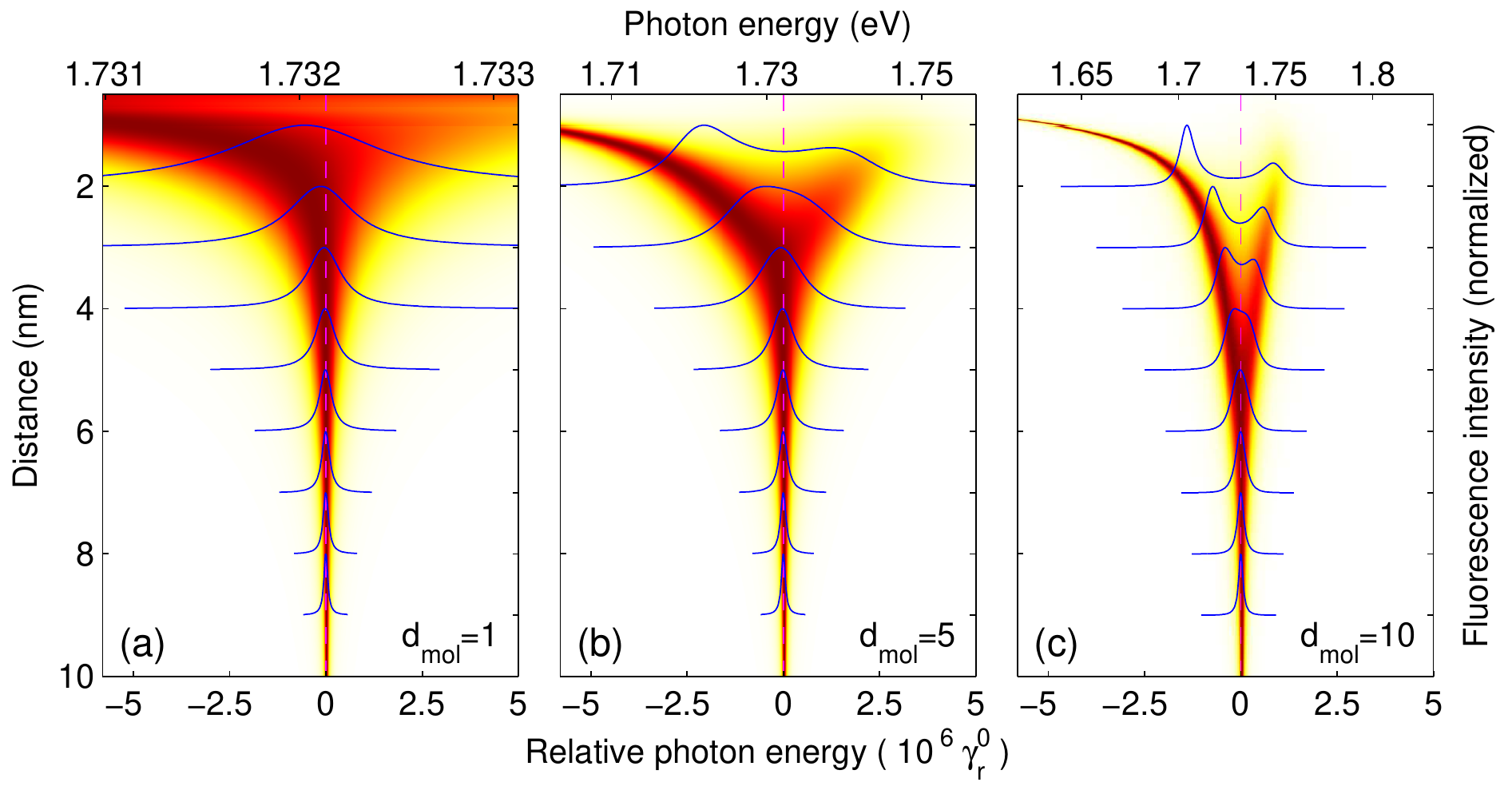}}
\caption{(Color online) Fluorescence spectra of the molecule in vicinity of the MNP for different molecular dipole moments, given in the panels in atomic units. The moments correspond to free-space decay rates $1/\gamma_r^0$ of approximately (a) 0.1 $\mu$s, (b) 4 ns, and (c) 1 ns. In the weak-coupling regime of panel (a) the line broadens with decreasing distance but does not split. In the strong-coupling regime of panels (b,c) the line splits at a given distance into two lines. }
\end{figure*}

\subsection{Molecule--MNP coupling}

With the SPP quantization we have now opened the quantum optics toolbox. This allows us to study strong coupling according to the standard prescription \cite{walls:95}. As for the description of the molecule we follow Refs.~\onlinecite{girard:95,girard:05b} who considered a generic few-level system. This approach is also best suited for other quantum emitters, such as collodial quantum dots. The inset of Fig.~1(b) shows the level scheme used in our calculations. It consists of the molecule ground state $0$ and two excited states $1$ and $2$. We assume that an external pump laser brings the molecule into the excited state $2$, where it decays non-radiativly with a given rate $\gamma_m$ to the optically active state $1$. This indirect process allows us to separate the excitation dynamics, which is not modified in presence of the MNP, from the relaxation dynamics of state $1$, which becomes strongly modified if the molecule and SPP are in resonance. The coherent part of the molecule-MNP dynamics is described by the Hamiltonian
\begin{equation}\label{eq:ham}
  H=H_{\rm mol}+H_{\rm pl}+H_{\rm mol-pl}+H_{\rm pump}\,.
\end{equation}
Here $H_{\rm mol}$ and $H_{\rm pl}$ describe the molecular states and the SPP modes, respectively, $H_{\rm mol-pl}$ is the coupling between the molecular dipole and the surface charge \eqref{eq:quantspp}, and $H_{\rm pump}$ is the interaction of the molecule with the pump laser. The last two terms in Eq.~\eqref{eq:ham} are described within the usual rotating-wave approximation.~\cite{walls:95} In addition, we account for the incoherent part of the dynamics through a master equation of Lindblad form \cite{walls:95,xu:04}
\begin{equation}\label{eq:master}
  \dot\rho=-i[H,\rho]-\mbox{$\frac 12$}\sum_i\left(L_i^\dagger L_i^{\phantom\dagger}\rho+
  L_i^\dagger L_i^{\phantom\dagger}\rho-2L_i^{\phantom\dagger}\rho L_i^\dagger\right)\,,
\end{equation}
where $\rho$ is the density matrix of the coupled molecule-SPP system. The Lindblad operators $L_i$ describe the various scattering channels of molecular decay, plasmon decay through Landau damping, and radiative decay \cite{remark.drude}.


\section{Results}\label{sec:results}

In our calculations we consider the cigar-shaped Ag MNP shown in Fig.~1. Other MNP shapes and metals will be discussed at the end. Figure~1(a) shows the spectra computed within our BEM approach~\cite{gerber.prb:07} for the Ag dielectric function of Ref.~\onlinecite{johnson:72} (solid line) and the Drude framework~\cite{remark.drude} (dashed line). Both spectra are in nice agreement, thus justifying the use of the Drude model. The energies of the SPP eigenmodes are indicated by triangles. Panel (b) reports the non-radiative and radiative  decay rates of the molecule as a function of molecule-MNP distance, which we compute in the weak-coupling regime according to the prescription of Refs.~\onlinecite{anger:06,gerber.prb:07}. One observes that the rates dramatically increase when the molecule approaches the MNP. Here the decay process becomes strongly altered by the nanoparticle, which acts as a supplemental antenna and converts part of the molecule's near field into radiation and Ohmic dissipation.

We next turn to the results of our master-equation approach. Figure~2 shows simulations based on the solution of Eq.~\eqref{eq:master} where the molecule is initially brought into the excited state $1$. Let us first consider the larger molecule-MNP distance of 8 nm (upper two lines). Through the coupling $H_{\rm mol-pl}$, the lowest SPP mode becomes populated and subsequently decays through Landau damping and radiation. After a transient at early times, both molecule and SPP population decay mono-exponentially with the same decay constant. In this regime the molecule drives the strongly damped plasmon mode and hereby constantly transfers energy to the MNP. Things change considerably when the molecule is brought closer to the MNP. For $z=4$ nm  (lower two lines) one observes a pronounced population beating between the molecule and the surface plasmon, superimposed on a sub-picosecond decay due to efficient plasmon damping. This beating behavior is a clear signature of strong coupling~\cite{walls:95} which occurs in a regime where the molecule-MNP coupling is stronger than the plasmon damping.

Although strong coupling is most apparent in the time domain, spectroscopy appears to be a more suitable tool for its experimental observation. We next turn to the study of the setup shown in the inset of Fig.~1(b), where a weak pump laser brings the molecule to the excited state $2$. This process is followed by an internal decay into the optically active state $1$ and a final relaxation to the groundstate. Again, the last process is strongly modified in presence of the MNP. In our calculations we use the master equation \eqref{eq:master} to compute the steady state solution. Once a stationary condition is reached, we can compute the fluorescence spectra from the Wiener-Khinchin theorem by means of the quantum regression theorem.~\cite{walls:95,xu:04} Figure 3 shows results of our simulations for three different molecular dipole moments. For the smallest dipole moment of panel (a), one observes that the line broadens when the molecule is brought closer to the MNP. We verified that the line broadening is precisely given by the sum of radiative $\gamma_r$ and non-radiative $\gamma_{nr}$ decay rates shown in Fig.~1(b). For the larger dipole moments investigated in panels (b,c), we observe that at a distance of a few nanometers the line splits, thus indicating the onset of strong coupling. Here excitation energy is coherently transfered between the molecule and the SPP.

In a generic model, where a quantum emitter is coupled to a single cavity mode, the polariton eigenmodes $\Omega_\pm$ of the coupled system are of the form~\cite{andreani:99}
\begin{equation}\label{eq:strongcoupling}
  \Omega_\pm+=\omega_0-
  \frac i4(\gamma_c+\gamma_m)\pm
  \sqrt{g^2-\left(\frac{\gamma_c-\gamma_m}4\right)^2}\,.
\end{equation}
Here, $\omega_0$ is the enery of the isolated molecule and cavity, which are assumed to be in resonance, $\gamma_m$ and $\gamma_c$ are the decay rates of the molecule and cavity, respectively, and $g$ is the coupling constant. Strong coupling occurs for $g>|\gamma_c-\gamma_m|/4$ and corresponds to the formation of a dressed state with finite lifetime. It is an intrinsic property of the coupling between the molecule and the cavity, and manifests itself as a doublet splitting of the emission lines. Quite generally, for the coupled molecule--MNP system Eq.~\eqref{eq:strongcoupling} is too simple, because the molecule couples not only to the MNP dipole mode but also to all other modes, and one must use a more refined description as we have done in this work. Nevertheless, Eq.~\eqref{eq:strongcoupling} allows us to estimate the pertinent parameters for strong coupling. From the plasmon decay rate $\gamma_0\sim 30$ fs for silver we can estimate a critical coupling strength of $g\approx\gamma_0/4\sim 5$ meV for the onset of strong coupling. Indeed, this value is in agreement with the results of Fig.~3 [as can be inferred, e.g., from the line broadening in panel (b) at the distance of 3 nm where the emission line starts to split]. For a coupling of the order of a few meV the approximation of a two-level system is justified for both molecules and quantum dots, although the true lineshape might be additionally influenced by internal degrees of freedom (e.g., vibrations) of the quantum emitter.

\begin{figure}
\centerline{\includegraphics[width=\columnwidth]{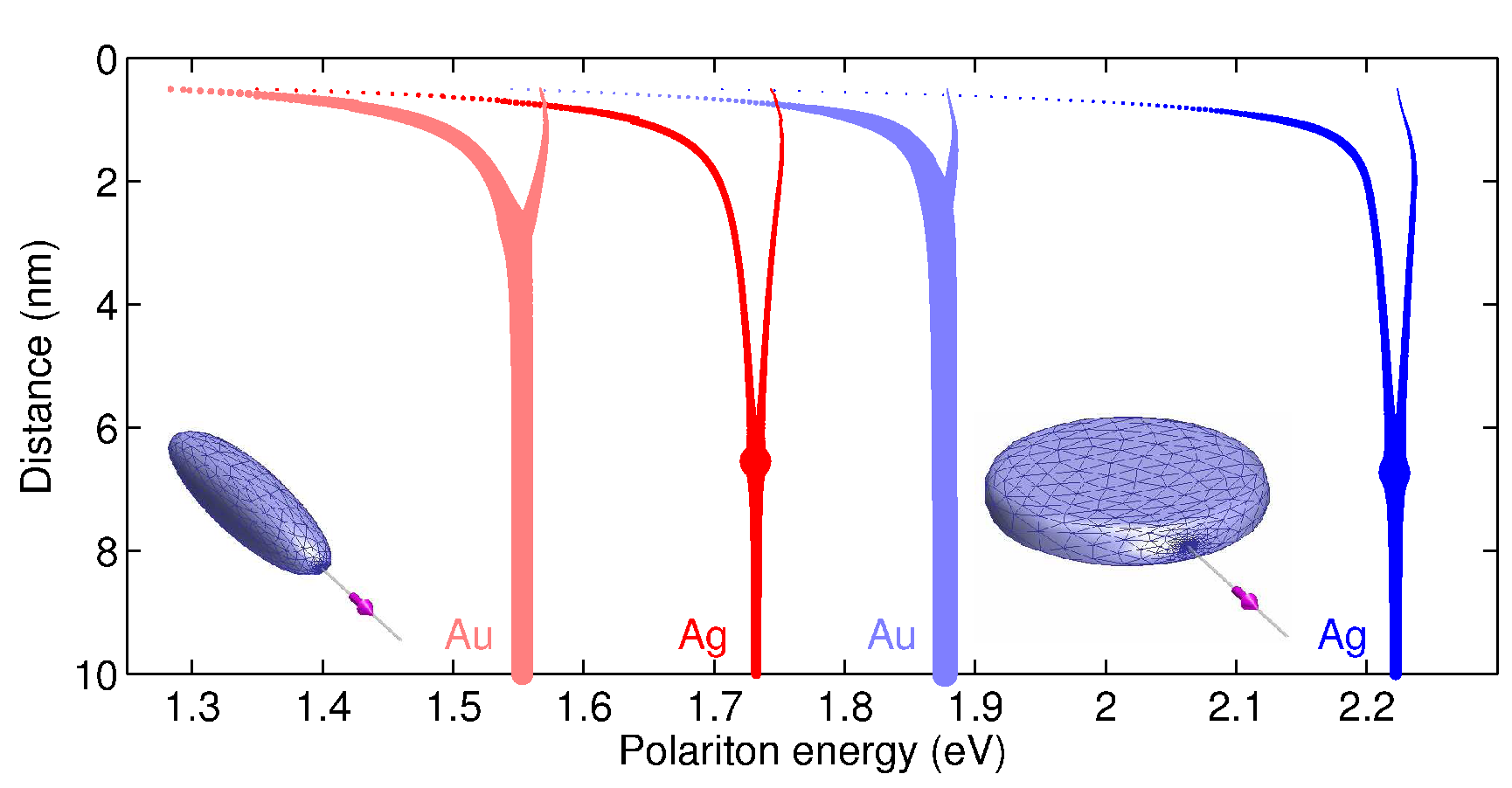}}
\caption{(Color online) Line positions and oscillator strengths of polariton modes for cigar-shaped (two lines on left-hand side) and disk-shaped (two lines on right-hand side) nanoparticles and Ag and Au. In Au the lines split at smaller distances because of the stronger plasmon damping.~\cite{remark.drude} We use $d=10$ atomic units.}
\end{figure}

When the molecule is brought even closer to the MNP, the oscillator strength of the high-energy line vanishes and the low-energy line becomes strongly red-shifted. In this regime, where the energy renormalization is of the order of several tens of meV, the description of the quantum emitter in terms of a generic few-level system is expected to break down. The strong redshift of the emission peaks is due to the attractive interaction between the molecule and the MNP, which is strongly enhanced at small distances, and the different oscillator strengths are associated to the different dipole moments of the predominantly MNP- and molecule-like polariton modes at lower and higher energy, respectively. We also found that moderate detunings between the molecule and SPP energies do not drastically change the behavior shown in the figures. Figure 4 shows that similar behavior is found for other MNP shapes and materials. We have also performed calculations for spherical nanoparticles. Unfortunately, for both silver and gold the resulting surface plasmons have energies in a spectral region where $d$-band scatterings set in,~\cite{ladstaedter.prb:04} and where the Drude description becomes questionable. Our results (not shown) indicate that for nanospheres strong coupling occurs at smaller distances than for the particle shapes shown in Fig.~4, which might be due to the larger number of plasmon modes to which the molecule can couple.~\cite{hohenester.ieee:08}


\section{Summary and Conclusions}\label{sec:conclusions}

In conclusion, we have studied strong coupling between a single molecule and a metallic nanoparticle within a fully quantum-mechanical approach. We have demonstrated that strong coupling is possible for realistic molecule and nanoparticle parameters, despite the strong plasmon damping, and should be observable in fluorescence spectroscopy through the splitting of emission peaks. Strong coupling is an important ingredient for future plasmonic-based quantum information schemes, and might play a significant role in biosensor applications.

\begin{acknowledgments}

We gratefully acknowledge most helpful discussions with Joachim Krenn and Alfred Leitner.

\end{acknowledgments}

\begin{appendix}

\section{}\label{app:supplementary}

In this appenix we derive Eq.~\eqref{eq:energyspp} and show how to quantize the SPP modes. Our starting point is the energy of a classical plasma, Eq.~\eqref{eq:energyplasma}. Let us first consider the first term on the right-hand side which describes the kinetic energy. From the relation $\bm v=-\nabla\Psi$ between the velocity field $\bm v$ and $\Psi$, we obtain for the continuity equation
\begin{equation}\label{eq:continuity}
  \partial_t n=-n_0\nabla\bm v=n_0\nabla^2\Psi\,,
\end{equation}
which gives us the relation between the density displacement $n$ and the velocity potential $\Psi$. In the following we consider surface charge distributions $\sigma$ which are nonzero only at the surface of the MNP. Integration of the continuity equation \eqref{eq:continuity} over a small cylinder $\Omega$ (height $h\to 0$ and base $\delta S$), which encloses a small surface element, then gives for the right-hand side of Eq.~\eqref{eq:continuity}
\begin{equation}\label{eq:intcontinuity}
  \int_\Omega n_0\nabla^2\Psi\,dV=\int_{\partial\Omega} n_0\,\hat{\bm n}\cdot\nabla\Psi\,dS
  \cong n_0\,\frac{\partial\Psi}{\partial\hat n}\,\delta S\,.
\end{equation}
Here, $\frac{\partial\Psi}{\partial\hat n}=\hat{\bm n}\cdot\nabla\Psi$ denotes the surface derivative of the velocity potential. Together with the left-hand side of Eq.~\eqref{eq:continuity} we find the link between $\Psi$ and $\sigma$,
\begin{equation}\label{eq:sigmapsi}
  n_0\,\frac{\partial\Psi}{\partial\hat n}=\dot\sigma\,.
\end{equation}
Using Green's first identity
%
%
we can rewrite the term for the kinetic energy in Eq.~\eqref{eq:energyplasma} as
\begin{equation}\label{eq:S5}
  \int_\Omega(\nabla\Psi)^2\,dV=
  \int_{\partial\Omega}\Psi\frac{\partial\Psi}{\partial\hat n}\,dS
  -\int_\Omega \Psi\nabla^2\Psi\,dV\,.
\end{equation}
As evident from the continuity equation \eqref{eq:intcontinuity}, for a pure surface charge distribution $\nabla^2\Psi$ is zero inside the metallic nanoparticle, and the second term on the right-hand side of \eqref{eq:S5} thus vanishes. We can now use the boundary-element method \cite{garcia:02,hohenester.prb:05,hohenester.ieee:08} to relate $\Psi$ to $\frac{\partial\Psi}{\partial\hat n}$. Our starting point is
\begin{equation}\label{eq:S6}
  \Psi(\bm r)=\int_{\partial\Omega}\left(
  G(\bm r,\bm s')\frac{\partial\Psi(\bm s')}{\partial\hat n}-
  \frac{\partial G(\bm r,\bm s')}{\partial\hat n}\Psi(\bm s')\right)\,\frac{dS'}{4\pi}\,,
\end{equation}
where $G(\bm r,\bm r')=1/|\bm r-\bm r'|$ is the free-space Green function. Performing the limit $\bm r\to\bm s$ in Eq.~\eqref{eq:S6} according to the prescription given in Refs.~\onlinecite{garcia:02,hohenester.prb:05,hohenester.ieee:08} and using the same notation as in these references, we obtain 
\begin{equation}\label{eq:S7}
  2\pi\Psi=G\Psi'-F\Psi\,.
\end{equation}
Here $\Psi'$ and $F$ are the surface derivatives of the velocity potential $\Psi$ and the Green function $G$, respectively, and $G$, $F$ and $\Psi$, $\Psi'$ are assumed to be convoluted in space. 

At this point it is convenient to switch to the boundary elements of the discretized MNP surface (see also inset of Fig.~1): $\Psi$ and $\sigma$ are vectors of the length of the number of surface elements, and $G$ and $F$ are matrices connecting the different surface elements. We can thus solve Eq.~\eqref{eq:S7} through inversion $\Psi=(2\pi+F)^{-1}G\,\Psi'$. Together with the relation \eqref{eq:sigmapsi}, the term for the kinetic energy can then be brought into the final form
\begin{equation}\label{eq:S8}
  \mbox{$\frac 12$}n_0\int_{\partial\Omega}\Psi\frac{\partial\Psi}{\partial\hat n}\,dS
  \longrightarrow \mbox{$\frac 12$}n_0\,\dot\sigma^T(2\pi+F)^{-1}G\,\dot\sigma\,.
\end{equation}
Here $\sigma^T$ denotes the transposed surface charge vector.

For the potential energy of Eq.~\eqref{eq:energyplasma} we follow the procedure given in Ref.~\onlinecite{hohenester.prb:05}. We start from a relation similar to Eq.~\eqref{eq:S6} but the velocity potential $\Psi$ replaced by the electrostatic potential $\Phi$. Taking the surface derivative inside and outside the MNP, we obtain
\begin{eqnarray*}
  (2\pi+F)\Phi&=&\phantom{-}G\,\Phi_1'\\
  (2\pi-F)\Phi&=& -G\,\Phi_2'\,.
\end{eqnarray*}
Here $\Phi_1'$ and $\Phi_2'$ denote the surface derivatives of the potential inside and outside the MNP. Multiplying the first equation with the dielectric constant $\epsilon_0$ of the metal and the second one with the dielectric constant $\epsilon_b$ of the embedding medium, gives after substraction
\begin{equation}\label{eq:S9}
  \bigl(2\pi(\epsilon_0+\epsilon_b)+(\epsilon_0-\epsilon_b)F\bigr)\,\Phi=
  G(\epsilon_0\Phi_1'-\epsilon_b\Phi_2')=4\pi\,G\sigma\,.
\end{equation}
To arrive at the last term of \eqref{eq:S9} we have used the boundary condition $\hat{\bm n}\cdot(\bm D_2-\bm D_1)=-\epsilon_b \Phi_2'+\epsilon_0\Phi_1'=4\pi\sigma$ of Maxwell's equation. Putting all above results together, we finally get Eq.~\eqref{eq:energyspp}.

To obtain the eigenmodes we first rewrite Eq.~\eqref{eq:energyspp} in the short-hand notation
\begin{equation}\label{eq:S11}
  H=\frac 1{2n_0}\left(\dot\sigma^T B\,\dot\sigma+\sigma^T A\,\sigma\right)\,,
\end{equation}
where the explicit form of the matrices $A$ and $B$ can be inferred from Eq.~\eqref{eq:energyspp}. The matrices $A$ and $B$ are symmetric and thus can be diagonalized simultaneously. Let $\omega_\lambda^2$ and $u_\lambda$ denote the eigenvalues and eigenvectors of the generalized eigenvalue problem
\begin{equation}\label{eq:S12}
  A\,u_\lambda=\omega_\lambda^2\,B\,u_\lambda\,.
\end{equation}
The eigenvectors $u_\lambda$ can be chosen real, and are orthogonal in the sense
\begin{equation}\label{eq:S13}
  u_\lambda^T\,B\,u_{\lambda'}=\beta_\lambda\,\delta_{\lambda\lambda'}\,.
\end{equation}
We can thus expand the surface charge distribution in terms of these eigenfunctions viz.
\begin{equation}\label{eq:S14}
  \sigma=\sum_\lambda \gamma_\lambda e^{i\omega_\lambda t}a_\lambda\,u_\lambda\,.
\end{equation}
Here $\gamma_\lambda=\sqrt{{2n_0}/({\omega_\lambda\beta_\lambda})}$, and $a_\lambda$ is an expansion coefficient for the eigenmode $\lambda$. Inserting this expression into equation \eqref{eq:energyspp} and performing the standard quantization procedure via a canonical transformation,~\cite{ritchie:57,barton:79,arista:01} then brings us to the plasmon Hamiltonian in second-quantized form
\begin{equation}\label{eq:S15}
  H_{\rm pl}=\sum_\lambda\omega_\lambda\,a_\lambda^\dagger a_\lambda^{\phantom\dagger}\,,
\end{equation}
with $a_\lambda^\dagger$ being the creation operator for the plasmon mode $\lambda$.

\end{appendix}


\begin{thebibliography}{35}
\bibitem[{rem()}]{remark.drude}
\bibinfo{note}{In our Drude description electrons are assumed to move freely in
  a medium with background dielectric constant $\epsilon_0$. We use
  $\epsilon_0=3.3$ for Ag and $\epsilon_0=10$ for Au. The dielectric constant
  of the embedding medium is set to $\epsilon_b=2.25$ throughout. The electron
  density $n_0$ is chosen such that the plasmon energy is 9 eV. For the
  lifetime $1/\gamma_0$ of the SPP modes we use 30 fs for Ag and 10 fs for Au.
  In our master equation approach the Lindblad operators for inter-molecular
  relaxation, plasmon damping, and radiative damping are of the form
  $\sqrt{\gamma_m}\,|1\rangle\langle 2|$, $\sqrt{\gamma_0}\,a_\lambda$, and
  $\sqrt{\gamma_r^0}\,(|0\rangle\langle 1|+[d_\lambda/d_m]a_\lambda)$,
  respectively. Here $\gamma_r^0$ is the radiative free-space decay rate of the
  molecule, and $d_m$ and $d_\lambda$ are the dipole moments of the molecule
  and the SPP modes, respectively.}
\end{thebibliography}

\end{document}